\documentclass[draftcls,onecolumn]{IEEEtran}

 \usepackage{cite}

\ifCLASSINFOpdf
\else
   \usepackage[dvips]{graphicx}
   \usepackage{epstopdf}
\fi
\usepackage{tikz}
\usetikzlibrary{arrows, fit}
\usepackage{verbatim}
\usepackage{amssymb}

\newcommand{\trace}{\operatorname{tr}}
\usepackage{algorithm}
\usepackage{amsthm}
\usepackage{amsmath}
\usepackage{hyperref}
\usepackage{cleveref}
\usepackage{lipsum}
\begin{document}
%
\title{Information Loss in the Human Auditory System}
%
%
%

\author{Mohsen~Zareian~Jahromi, Adel~Zahedi,
        Jesper~Jensen,
        and~Jan~\O stergaard.
\thanks{M.Z. Jahromi, A. Zahedi, and J. \O stergaard  are with the Department of Electronic Systems, Aalborg University, Aalborg 9220, Denmark (e-mail: \{mzj,adz,jo\}@es.aau.dk). }
\thanks{J. Jensen is with the Department of Electronic Systems, Aalborg University, Aalborg 9220, Denmark, and also with Oticon A/S, Sm\o rum 2765, Denmark
(e-mail: jje@es.aau.dk; jesj@oticon.com).}
\thanks{The work has received funding from VILLUM FONDEN Young Investigator Programme, under grant agreement No. 10095.}
}

\maketitle

\begin{abstract}
	
	From the eardrum to the auditory cortex, where acoustic stimuli are decoded, there are several stages of auditory processing and transmission where information may potentially get lost. In this paper, we aim at quantifying the information loss in the human auditory system by using information theoretic tools. 
	To do so, we consider a speech communication model, where words are uttered and sent through a noisy channel, and then received and processed by a human listener. 
	We define a notion of information loss that is related to the human word recognition rate. To assess the word recognition rate of humans, we conduct a closed-vocabulary intelligibility test. We derive upper and lower bounds on the information loss. Simulations reveal that the bounds are tight and we observe that the information loss in the human auditory system increases as the signal to noise ratio (SNR) decreases. Our framework also allows us to study whether humans are optimal in terms of speech perception in a noisy environment. Towards that end, we derive optimal classifiers and compare the human and machine performance in terms of information loss and word recognition rate. We observe a higher information loss and lower word recognition rate for humans compared to the optimal classifiers. In fact, depending on the SNR, the machine classifier may outperform humans by as much as 8 dB. This implies that for the speech-in-stationary-noise setup considered here, the human auditory system is sub-optimal for recognizing noisy words.
	
\end{abstract}


%
\IEEEpeerreviewmaketitle

\section{Introduction}
As an acoustic signal enters the ear, it passes several processing stages until the information it carries is decoded in the brain. There exist numerous works that have studied and modeled some stages of auditory processing, from biophysical to computational models \cite{seneff1990joint,ghitza1986auditory,lyon1982computational,tchorz1999model,dau1996quantitative,yang1992auditory,huettel1999using,carney1993model,siebert1968stimulus,patterson1995time,harczos2013making}. Due to the information processing and transmission at each stage, some information loss may occur. In this study, our motivation is to quantify the information loss in the human auditory system, from the eardrum to the speech decoding stage in the brain. This is a first step towards assessing the information loss of the individual components in the human auditory system. We model a speech communication system, where a speaker utters a word from a fixed dictionary and the word waveform passes a noisy communication channel, before it is classified by a human listener.


Our key idea is to define a notion of information loss, which is related to the number of words that are not correctly recognized. Since a certain degree of information is also lost in the acoustic communication channel, we normalize the information loss so it describes the ratio of the amount of information lost due to being processed by the listener and the total amount of information that reaches the listener's eardrum. 

To assess the word recognition rate of humans, we conduct a closed-vocabulary intelligibility test. This test is a listening test that reflects key properties of the DANTALE II intelligibility test \cite{wagener2003design}, where intelligibility is determined by presenting speech stimuli contaminated by noise to test subjects, and calculating the word recognition rate.

We quantify the information flow through the acoustic channel by establishing computable lower and upper bounds on the mutual information between the words being uttered and the output of the noisy acoustic channel. 
Simulations reveal that the bounds are tight, and we observe that the information loss in the human auditory system increases as the signal to noise ratio (SNR) decreases. We also observe, that the information loss has an inverse relationship with the word recognition rate of humans.

Our framework further allows us to assess whether humans are optimal in terms of speech perception in a noisy environment. It may be hypothesized that the ability to understand speech under varying acoustic conditions has provided humans with an evolutionary advantage. In particular, it has been hypothesized that animals are close-to-optimal at performing tasks that are important for their survival, e.g. they transfer information optimally from the sensory world to the brain \cite{bialek1991optimal, atick1992could}. Rieke et al. \cite{rieke1995naturalistic} studied the peripheral auditory system of the bullfrog. They first estimated the input stimulus (i.e., stimulus reconstruction) by linearly filtering the output (spike trains) of the auditory system. They then measured the information rate carried by spike trains, e.g. the rate at which the spike trains remove uncertainty about the sensory output. This information rate reaches its upper bound, i.e. the stimulus is transfered optimally, when the input stimulus is a natural sound, rather than a synthetic stimulus. This indicates that the auditory system of this organism is tuned to natural stimuli. Similar studies have been done on different organisms, e.g. \cite{bialek1993bits, bialek1991reading}. For example, in \cite{bialek1991reading}, they investigated a single neuron, which is sensitive to movements in the visual system of the blowfly in terms of information rate and conclude that the visual system of this organism transmits information optimally.

To answer the question of whether humans are optimal in terms of speech perception, we derive optimal classifiers and compare human and machine performances in terms of their information losses and word recognition rates. 
It is observed that with equal SNR, humans have higher information loss compared to the optimal classifiers. In fact, depending on the SNR, the machine classifier may outperform humans by as much as 8 dB, depending on the prior knowledge assumed available to the classifier. This implies that for the speech-in-stationary-noise setup considered here, the human auditory system is sub-optimal for recognizing noisy word.

\subsection{Overview of the Paper}
The rest of the paper is organized as follows. In Sec.\!\! II, we
describe our speech communication model. In Sec.\!\! III, we introduce and quantify the relative information loss in the speech communication model and find lower and upper bounds for it. We also derive the optimal classifiers in this section. We explain the simulation study and report the results in Sec.\!\! IV and discuss them in Sec.\!\! V. Sec.\!\! VI concludes the paper.

\subsection{Notation}
We denote random vectors and random scalars with boldface uppercase, and italic uppercase letters  respectively.  Boldface lowercase  and italic lowercase  letters are used for denoting deterministic vectors and deterministic scalars respectively. We denote the expectation operation with respect to random variable $ \textbf{y} $ by $ E_{\textbf{y}}[.] $. The information theoretic quantities of differential entropy,  entropy, and mutual information are denoted by $ h(.)  $, $ H(.) $ and $ I(. ; .) $, respectively. The trace operation and the matrix determinant are denoted by $ tr(.) $ and $ |.| $ respectively. We denote Markov chains by
two-headed arrows; e.g $ X\leftrightarrow Y\leftrightarrow Z $. The probability mass function (PMF) is denoted by $ P(.) $,  and $ f (.) $ is used for the probability density function (PDF). For example, $ f_{\textbf{Y}|m}(\textbf{y}|m)  $ denotes the conditional PDF of $ \textbf{Y} $ given $ M=m $. The notation $ \textbf{y}^{z} $ represents the sequence $ [\textbf{y}_{1},\textbf{y}_{2},...,\textbf{y}_{z}] $.
%
%
%
%

\section{Communication Model}
\label{sec:Model_Classifi}
\tikzstyle{int}=[draw, rectangle, fill=blue!20, minimum size=4em, text centered, text width=2cm]
\tikzstyle{block} = [draw,fill=blue!20,minimum size=2em]
\tikzset{blue dotted/.style={draw=blue!50!white, line width=1pt,dash pattern=on 1pt off 4pt on 6pt off 4pt, inner sep=2mm, rectangle, rounded corners}}
\begin{figure}[!]
	\centering %
	\begin{tikzpicture}[node distance=2cm,auto,>=latex']
\node at (2,1) (a) {$\textbf{X}_{1}$};
\node at (2,0) (a1) {$\textbf{X}_{2}$};
\node at (2,-1) (a2) {.};
\node at (2,-1.2) (a3) {.};
\node at (2,-1.4) (a4) {.};
\node at (1.9,-2.2) (a5) {$\textbf{X}_{\Gamma}$};
\node at (3.2,-1) (a6) {};
\draw[->] (a.east) -- +(0.5,0); 
\draw[->] (a1.east) -- +(0.5,0); 
\draw[->] (a5.east) -- +(0.5,0); 
\path[<->,bend right=45] (3.4,-.0) edge node {} (2.8,-.3) ;

	\node  [int, right of=a2,node distance=3.2cm] (b) {Noisy Acoustic channel};
	\draw[->] (a6) edge node {$ \textbf{X}_{m} $} (b) ; 
	\node  (y)  at ([shift=({1.5cm,.2cm})]b)  {$\textbf{Y}$};
	\node  (c)  [int, right of=b,node distance=3cm]{Listener/\\Classifier};
	\node (end)  [right of=c, node distance=1.8cm] {$m^{*}$};
\path[ <- ] (2.8,1)  edge node {}  (3.33,-1.0);
	\path[->] (b) edge node {} (c);
	\draw[->] (c) edge node {} (end) ;
	\end{tikzpicture}
	\caption{Block diagram of the speech communication model}
		\label{A_General_Block}
\end{figure}
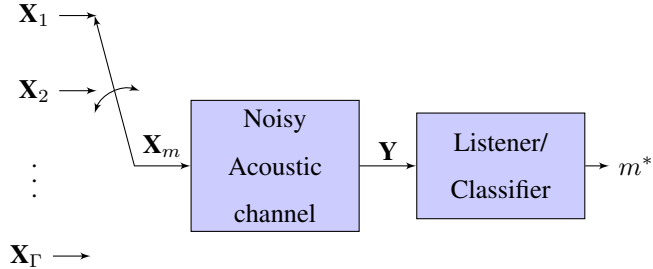
\label{subsec:DANTALE-parad}
 Figure \ref{A_General_Block} illustrates the speech communication model that is composed of three parts: Speaker, Noisy Channel, and Listener/Classifier. We elaborate on these parts below.
 
 \subsection{Speaker}
 The speaker constructs sequences of words by choosing words randomly  from fixed dictionaries that are also known by the listener/classifier. Let us consider a set $ c=\{1,2,...,\Gamma\} $. The waveform of the $ m^{th} $ word is modeled as a random vector $ \textbf{X}_{m} \in \mathbb{R}^{n}, m \in c $  that contains $ n $ samples. The discrete random variable $ M $ indexes the word that is picked and uttered. $ P(M=m) $ denotes the probability that the $ m^{th} $ word is chosen.  
 \subsection{Noisy Channel}
 \begin{figure}[!]
 	\centering %
 	\tikzstyle{int}=[draw, fill=blue!20, minimum size=4em,text centered, text width=2cm]
 	\tikzstyle{init} = [pin edge={to-,thin,black}]
 	\tikzstyle{sum} = [draw, fill=blue!20, circle, node distance=2cm]
 	\tikzset{blue dotted/.style={draw=blue!50!white, line width=1pt,dash pattern=on 1pt off 4pt on 6pt off 4pt, inner sep=.7mm, rectangle, rounded corners}}
 	\tikzset{black dotted/.style={draw=black!50!white, line width=1pt, inner sep=1mm, rectangle, rounded corners}}
 	
 	\begin{tikzpicture}[node distance=2.5cm,auto,>=latex']

 	\node (b) [sum, pin={[init]below:},][node distance=1.5cm, right of=a6]  {$\times $};
 	\draw[->] (a6) edge node {$ \textbf{X}_{m} $} (b) ; 
 	\node (c) [sum, pin={[init]above:},][node distance=1.2cm, right of=b]  {$+$};
 	\node  (z)  at ([shift=({0cm,-.9cm})]b)  {$\sqrt{\theta}$};
 	\node  (n)  at ([shift=({0cm,.9cm})]c)  {$\textbf{W}$};
 	\node  (y)  at ([shift=({.7cm,.2cm})]c)  {$\textbf{Y}$};
 	
 	
 	\draw[->] (c.east) -- +(0.5,0); 
 	\path[->] (b) edge node {} (c)  ;
 	\end{tikzpicture}
 	\caption{The word $ \textbf{X}_{m} $ is conveyed over a noisy acoustic channel.}
 	\label{Dantale_Block}
 \end{figure}
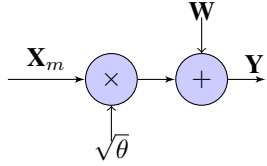
 The noisy channel is composed of a clean speech term multiplied by a scaling factor and an additive noise term (Fig. \ref{Dantale_Block}). The additive noise, $ \textbf{W} $, is zero-mean coloured Gaussian $ \textbf{W}\sim \mathcal{N} (\textbf{0},\mathbf{\Sigma}_{\textbf{W}}) $ and has a long-term spectrum similar to the average long-term spectrum of the clean words. The scale factor $ \sqrt{\theta} $  serves to modify the SNR, which is defined as the ratio of the average power of the words $\textit{p}_{ave}=E_{m}[\|\sqrt{\theta}\textbf{x}_{m}\|_{2}^{2}]$, to the noise power $ p_{noise}=E[ \|\textbf{W}\|_{2}^{2} ]  $. Without loss of generality, we fix $ p_{noise}=E[ \|\textbf{W}\|_{2}^{2} ]  =E_{m}[\|\sqrt{\theta}\textbf{x}_{m}\|_{2}^{2}]$, then $ \text{SNR}=\frac{\textit{p}_{ave}}{\textit{p}_{noise}}= \theta$.  The received  word waveform $ \textbf{Y} $ is  expressed as:
 \begin{align}
 \label{Main_Equation}
 \textbf{Y}=\sqrt{\theta} \textbf{X}_{m}+ \textbf{W}.  
 \end{align}
 \subsection{Listener/Classifier}
     The listener receives the noisy word waveform $ \textbf{y} $ and attempts to recognize it by mapping it to one of the words in the dictionary. The random variable $ M^{*} $ specifies the word selected by the listener/classifier. 
 \section{Analysis}
 \subsection{Relative Information Loss}
 Consider the speech communication model in Fig.\!\! \ref{A_General_Block}. Since $ m^{*} $ is a deterministic function of $ \textbf{y} $,  we can write:
  \begin{align}
 \label{Markov_chain_requirment}
 P(M^{*}=m^{*}|\textbf{y},M=m)=P(M^{*}=m^{*}|\textbf{y}).
 \end{align}
 Equation \eqref{Markov_chain_requirment} implies that $ M,\textbf{Y} $ and $ M^{*} $ form a Markov chain,  $ M \leftrightarrow \textbf{Y} \leftrightarrow  M^{*} $, from which, by the data processing inequality, we have\cite{cover2012elements}:
 \begin{align}
 \label{Markov_chain}
 0\leq I(M;M^{*}) \leq  I(M;\textbf{Y}) \leq H(M).
 \end{align}
  As mentioned above, information may be lost at any stage of auditory processing. As the result, the amount of information common between $ M^{*} $ and $ M $, i.e.  $ I(M;M^{*}) $, is less than $ I(M;\textbf{Y}) $. Therefore, the difference between $ I(M;\textbf{Y}) $ and $ I(M;M^{*}) $ is the amount of information that is lost in the listener/classifier part (cf. Fig \ref{A_General_Block}). Based on this argument, we define the information loss, $ l\geq 0 $,  as follows:
  \begin{align}
  \label{Loss_Genral}
 l \triangleq I(M;\textbf{Y})-I(M;M^{*}).
 \end{align}
 
   If the classifier block in Fig.\!\! \ref{A_General_Block} is performed by human listeners,  $ l $ quantifies the  amount of  information that is lost in the human auditory system from the eardrum to the decoding stage in the brain. However, the information loss $ l $ in \eqref{Loss_Genral} does not reveal the size of the loss  compared to the total information that reaches the eardrum $ I(M;\textbf{Y}) $. Thus,  we  introduce a relative information loss $ l_{I} $ as:  
 \begin{align}
 l_{I}=\dfrac{I(M;\textbf{Y})-I(M;M^{*})}{I(M;\textbf{Y})}=1-\dfrac{I(M;M^{*})}{I(M;\textbf{Y})}.
 \end{align}
  When the decoding block is performed by human listeners, $ l_{I} $  can be interpreted as the fraction of the information reaching the eardrum, which is actually used for decoding the speech signal. From \eqref{Markov_chain}, it is easy to show that $ 0\leq l_{I}\leq 1 $. 
\subsection{Bounds on  $ I(M;\textbf{Y}) $}
 To calculate $ l_{I} $, let us start with the definition of $ I(M;\textbf{Y}) $:
 \begin{align}
 \label{Main_MI}
 I(M;\textbf{Y})=h(\textbf{Y})-h(\textbf{Y}|M).
 \end{align}
  The PDF of $ \textbf{Y} $ can be obtained as: 
 \begin{align}
 \label{Mixture_Y}
 f_{\textbf{Y}}(\textbf{y})&=\sum_{m=1}^{\Gamma}P(M=m)f_{\textbf{Y}|m}(\textbf{y}|m).
 \end{align}
  As  seen, $ f_{\textbf{Y}}(\textbf{y}) $ is a mixture of distributions, and to the best of the authors knowledge, there exists no closed-form expression for the entropy of a mixture distribution. However, we can find upper and lower bounds for  $ I(M;\textbf{Y}) $, and, consequently, for $l_{I} $.  
  
  We first  divide $ \textbf{X}_{m} $ and $  \textbf{Y} $ into successive, non-overlapping frames each of   length  $ k $: 
  \begin{align}
  \label{Successive_Frames}
  \textbf{X}_{m}&=[\textbf{X}_{m,1},\textbf{X}_{m,2},...,\textbf{X}_{m,v}],\nonumber\\\textbf{Y}&=[\textbf{Y}_{1},\textbf{Y}_{2},...,\textbf{Y}_{v}],\nonumber\\
  \end{align}
  where $ \textbf{X}_{m,z} \in \mathbb{R}^{k}, z \in \{ 1,2,...,v\}$  is the $ z^{th} $ frame of the $ m^{th} $ word, $ \textbf{Y}_{z} \in \mathbb{R}^{k} $ is the $ z^{th} $ frame of the noisy word, and $ v $ denotes the number of frames. Using the product rule, we can write:
  \begin{align}
  \label{chain_rule}
f_{\textbf{Y}}(\textbf{y})=\prod_{z=1}^{v}f_{\textbf{Y}_{z}|\textbf{y}^{z-1}}(\textbf{y}_{z}|\textbf{y}^{z-1}).
  \end{align}
Note that we do not assume frames to be independent.

Let $ I_{l}(M;\textbf{Y}) $ and $ I_{u}(M;\textbf{Y}) $ denote lower and upper bounds for $ I(M;\textbf{Y}) $, and let $ D_{KL} $ and $ C_{\beta} $ ($ \beta\in [0,1] $) denote the KL-divergence and the Chernoff $ \beta $-divergence between two distributions respectively \cite{renyi1961measures}:
\begin{align}
\label{KL_C_Difinitions}
&D_{KL}(g||f)=\int g(x)\log\dfrac{g(x)}{f(x)}dx,\nonumber\\
&C_{\beta}(g||f)=-\log\int g^{\beta}(x)f^{1-\beta}(x)dx.
\end{align} 

\renewcommand{\labelenumi}{(\roman{enumi})}
\renewcommand{\theenumi}{(\roman{enumi})}
\newtheorem{lemma}{Lemma}
\begin{lemma}
	\label{lemma_Up_Low_In_General}
	 The mutual information between  $ M $ and  $ \textbf{Y} $ is lower and upper bounded by:
	\begin{align}
	\label{I_U_L_Extended_Lemma_General}
	&I_{l}(M;\textbf{Y})=\max\left\{0,-\sum_{m=1}^{\Gamma}P(M=m)\right. \nonumber\\&\left.  \times\log\left(\sum_{m'=1}^{\Gamma}P(M=m')\right.\right.\nonumber\\& \left. \left. \times \exp\left(-C_{\beta}(f_{\textbf{Y}^{v-1}|m}f_{\textbf{Y}_{v}|\textbf{y}^{v-1},m}||f_{\textbf{Y}^{v-1}|m'}f_{\textbf{Y}_{v}|\textbf{y}^{v-1},m'})\right)\right)\right\}\nonumber.
	\end{align}
	and
	\begin{align}
	&I_{u}(M;\textbf{Y})=\min\left\{H(M),-\sum_{m=1}^{\Gamma}P(M=m)\right. \nonumber\\&\left. \times\log\left(\sum_{m'=1}^{\Gamma}P(M=m')\right.\right.\nonumber\\&\left. \left. \times  \exp\left(-\sum_{z=1}^{v}E_{\textbf{y}^{z-1}}\left[D_{KL}(f_{\textbf{Y}_{z}|\textbf{y}^{z-1},m}||f_{\textbf{Y}_{z}|\textbf{y}^{z-1},m'})\right]\right)\right)\right\}.
	\end{align}

	\begin{proof}
		See Appendix A.
	\end{proof} 
\end{lemma}

\subsubsection{Gaussian Case}
As seen from \eqref{I_U_L_Extended_Lemma_General}, the lower and upper bounds for $ I(M;\textbf{Y}) $ depend on the PDF of $f_{\textbf{Y}_{z}|\textbf{y}^{z-1},m}(\textbf{y}_{z}|\textbf{y}^{z-1},m) $. From \eqref{Main_Equation}, we have: 
\begin{align}
\label{Y_z}
\textbf{Y}_{z}=\sqrt{\theta} \textbf{X}_{m,z}+ \textbf{W}_{z}.  
\end{align}
At high and medium SNRs where humans successfully recognize all the words, the information loss is zero. We thus focus on low SNRs in this work. At low SNRs ($ \theta\ll 1 $), the additive Gaussian noise $ \textbf{W}_{z} $ in \eqref{Y_z} is dominant. Therefore, it is reasonable to assume that  $f_{\textbf{Y}_{z}|m}(\textbf{y}_{z}|m) $ approximately follows a  Gaussian distribution $f_{\textbf{Y}_{z}|m}(\textbf{y}_{z}|m)\sim \mathcal{N} (\textbf{0},\mathbf{\Sigma}_{\textbf{Y}_{z}|m})$. Consequently, $f_{\textbf{Y}_{z}|\textbf{y}^{z-1},m}(\textbf{y}_{z}|\textbf{y}^{z-1},m) $ also follows a  Gaussian distribution,  $f_{\textbf{Y}_{z}|\textbf{y}^{z-1},m}(\textbf{y}_{z}|\textbf{y}^{z-1},m)\sim \mathcal{N} (\mathbf{\mu}_{\textbf{Y}_{z}|\textbf{y}^{z-1},m},\mathbf{\Sigma}_{\textbf{Y}_{z}|\textbf{y}^{z-1},m})$, where:
\begin{align}
\label{Covariance_Matrix}
&\mathbf{\Sigma}_{\textbf{Y}_{z}|\textbf{y}^{z-1},m}=\mathbf{\Sigma}_{\textbf{Y}_{z}|m}-\mathbf{\Sigma}_{\textbf{Y}_{z}\textbf{Y}^{z-1}|m}\mathbf{\Sigma}_{\textbf{Y}^{z-1}|m}^{-1}(\mathbf{\Sigma}_{\textbf{Y}_{z}\textbf{Y}^{z-1}|m})^{T},\nonumber\\&\mathbf{\Sigma}_{\textbf{Y}_{z}\textbf{Y}^{z-1}|m}=\theta\mathbf{\Sigma}_{\textbf{X}_{m,z}\textbf{X}_{m}^{z-1}}+\mathbf{\Sigma}_{\textbf{W}_{z}\textbf{W}^{z-1}},\nonumber\\&\mathbf{\Sigma}_{\textbf{Y}^{z-1}|m}=\theta\mathbf{\Sigma}_{\textbf{X}_{m}^{z-1}}+\mathbf{\Sigma}_{\textbf{W}^{z-1}},\nonumber\\&\mathbf{\mu}_{\textbf{Y}_{z}|\textbf{y}^{z-1},m}=\mathbf{\Sigma}_{\textbf{Y}_{z}\textbf{Y}^{z-1}|m}\mathbf{\Sigma}_{\textbf{Y}^{z-1}|m}^{-1}\textbf{y}^{z-1}.
\end{align}

 When $ \textbf{Y}_{z} $ is Gaussian (i.e. for low SNRs), we obtain the closed form expression for the KL-divergence and the Chernoff $ \beta $-divergence of two Gaussians \cite{renyi1961measures} in \eqref{I_U_L_Extended_Lemma_General}:

\begin{align}
\label{Distances_Gaussians}
&E_{\textbf{y}^{z-1}}\left[D_{KL}(f_{\textbf{Y}_{z}|\textbf{y}^{z-1},m}||f_{\textbf{Y}_{z}|\textbf{y}^{z-1},m'})\right]=\nonumber\\&\dfrac{1}{2}\left[\log\dfrac{|\mathbf{\Sigma}_{\textbf{Y}_{z}|\textbf{y}^{z-1},m'}|}{|\mathbf{\Sigma}_{\textbf{Y}_{z}|\textbf{y}^{z-1},m}|}+\trace(A^{T}\mathbf{\Sigma}_{\textbf{Y}^{z-1}|m})-\right.\nonumber\\&\left. \trace((\mathbf{\Sigma}_{\textbf{Y}_{z}|\textbf{y}^{z-1},m'})^{-1}(\mathbf{\Sigma}_{\textbf{Y}_{z}|\textbf{y}^{z-1},m}))-k\right],\nonumber\\&C_{\beta}(f_{\textbf{Y}^{v-1}|m}f_{\textbf{Y}_{v}|\textbf{y}^{v-1},m}||f_{\textbf{Y}^{v-1}|m'}f_{\textbf{Y}_{v}|\textbf{y}^{v-1},m'})=\nonumber\\&\dfrac{1}{2}\log\left(|\mathbf{\Sigma}_{\textbf{Y}_{v}|\textbf{y}^{v-1},m}|^{\beta}|\mathbf{\Sigma}_{\textbf{Y}_{v}|\textbf{y}^{v-1},m'}|^{1-\beta}\nonumber\right.\\&\left.\times|\beta\mathbf{\Sigma}_{\textbf{Y}_{v}|\textbf{y}^{v-1},m}+(1-\beta)\mathbf{\Sigma}_{\textbf{Y}_{v}|\textbf{y}^{v-1},m'}||\mathbf{\Sigma}_{\textbf{Y}_{v-1}|m}|^{\beta}\nonumber\right.\\&\left.|\mathbf{\Sigma}_{\textbf{Y}_{v-1}|m'}|^{1-\beta}|\beta\mathbf{\Sigma}_{\textbf{Y}_{v}|m}+(1-\beta)\mathbf{\Sigma}_{\textbf{Y}_{v}|m'}+B|\right),
\end{align}
where 
\begin{align}
&A=(\mathbf{\Sigma}_{\textbf{Y}_{z}\textbf{Y}^{z-1}|m}\mathbf{\Sigma}_{\textbf{Y}^{z-1}|m}^{-1}-\mathbf{\Sigma}_{\textbf{Y}_{z}\textbf{Y}^{z-1}|m'}\mathbf{\Sigma}_{\textbf{Y}^{z-1}|m'}^{-1})^{T}\nonumber\\&\times\mathbf{\Sigma}_{\textbf{Y}_{z}|\textbf{y}^{z-1},m'}(\mathbf{\Sigma}_{\textbf{Y}_{z}\textbf{Y}^{z-1}|m}\mathbf{\Sigma}_{\textbf{Y}^{z-1}|m}^{-1}-\mathbf{\Sigma}_{\textbf{Y}_{z}\textbf{Y}^{z-1}|m'}\mathbf{\Sigma}_{\textbf{Y}^{z-1}|m'}^{-1})\nonumber\\&B=\beta(1-\beta)(\mathbf{\Sigma}_{\textbf{Y}_{z}\textbf{Y}^{z-1}|m}\mathbf{\Sigma}_{\textbf{Y}^{z-1}|m}^{-1}-\mathbf{\Sigma}_{\textbf{Y}_{z}\textbf{Y}^{z-1}|m'}\mathbf{\Sigma}_{\textbf{Y}^{z-1}|m'}^{-1})^{T}\nonumber\\&\times(\beta\mathbf{\Sigma}_{\textbf{Y}_{v}|\textbf{y}^{v-1},m'}+(1-\beta)\mathbf{\Sigma}_{\textbf{Y}_{v}|\textbf{y}^{v-1},m})^{-1}\nonumber\\&\times(\mathbf{\Sigma}_{\textbf{Y}_{z}\textbf{Y}^{z-1}|m}\mathbf{\Sigma}_{\textbf{Y}^{z-1}|m}^{-1}-\mathbf{\Sigma}_{\textbf{Y}_{z}\textbf{Y}^{z-1}|m'}\mathbf{\Sigma}_{\textbf{Y}^{z-1}|m'}^{-1}).
\end{align}
We will  use this result for Gaussian signals  to bound the relative information loss in the next subsection and derive optimal classifiers in subsection III.D.
\subsection{Bounds on Relative Information Loss}
 Suppose that $ P_{c} \triangleq P(M^{*}=M) $ is the word recognition rate. From the definition of mutual information we have:
 
 \begin{align}
 \label{adel}
 I(M;M^{*})&=H(M)-H(M|M^{*})\nonumber\\
           &= \log(\Gamma)-P_{c}\log\left(\!\frac{1}{P_{c}}\!\right)\!-\!(1-P_{c})\log\left(\!\frac{\Gamma-1}{1-P_{c}}\!\right),
 \end{align} 
where the first term on the right-hand side of \eqref{adel} follows from the fact that $M$ is drawn uniformly from the set $\{1,2,\cdots,\Gamma\}$, and the last two terms result from calculating $-H(M|M^{*})$ using the definition of conditional entropy.

As can be seen from \eqref{adel},  $ I(M;M^{*}) $ depends on $ P_{c}$. In order to calculate this probability, if the classifier block is performed by human listeners, we perform a listening test (see section IV for more details). We also derive optimal classifiers in the next subsection, to obtain $ P_{c} $, when  the decoder part is performed by the optimal machine classifier. 

 Using \eqref{I_U_L_Extended_Lemma_General} and \eqref{adel},  the upper and lower bound for the relative information loss are obtained as follows:
\begin{align}
\label{Lower_Upper_Information_Loss}
	l_{I}^{u}&=1-\dfrac{I(M;M^{*})}{I_{u}(M;\textbf{Y})},\nonumber\\l_{I}^{l}&=\begin{cases}
1-\dfrac{I(M;M^{*})}{I_{l}(M;\textbf{Y})}, & \text{if $I(M;M^{*})<I_{l}(M;\textbf{Y})$}.\\
0, & \text{otherwise},
	\end{cases}
\end{align}
where $ l_{I}^{u} $ and $ l_{I}^{l} $ denote the upper and lower bound for the relative information loss, respectively. The tightness of the lower and upper bounds for $ l_{I} $ depends on how tight the lower and upper bounds for the entropy $ h(\textbf{Y}) $  of the Gaussian mixture model (GMM) are. In \cite{kolchinsky2017estimating}, it is shown that the upper and lower bounds for the entropy of the GMM are significantly tighter than well-known existing bounds \cite{huber2008entropy} \cite{jebara2003bhattacharyya}\cite{joe1989estimation}. We also observe that in our case, the upper and lower bounds for $ l_{I} $ are tight (See Section IV).
\subsection{Optimal Classifier}
In this section, we derive the optimal classifiers for our speech communication model.  Since the performance of the optimal classifiers will be compared to that of the humans, in order to have a fair comparison, we make an assumption and some requirements for the optimal classifiers based on the situations that human listeners encounter. 
\begin{enumerate}
	\item\label{as1} We assume that test subjects are able to learn and store a model of the words based on the spectral envelope contents of sub-words  encountered during the training phase.  In a similar manner,  subjects create an internal noise model. This assumption is inspired by \cite{schadler2015matrix,meyer2013comparison},  which suggest that humans  build internal statistical models of the words based on characteristics of the spectral contents of sub-words. In our classifier, this is achieved by allowing the classifier to have access  to training data in terms of average short-term speech spectra of the clean speech and of the noise.
	
\end{enumerate}
 In addition, we impose the following requirements on the subjective listening test and the classifier:
\begin{enumerate}
	\item\label{re1} We design  a classifier, which maximizes the probability of correct word detection. This reflects the fact that subjects are instructed to make a best guess of each noisy word.  
	
	\item\label{re2} When listening to the stimuli, i.e. the noisy sentences, the subjects are not informed about the SNRs a priori. In a similar manner, the classifier does not rely on a priori knowledge of the SNR.  Therefore from the decoder point of view, the realization of the SNR, $ \theta $, is  uniformly drawn from  $ \Theta \sim \mathcal{U}(a, b) $ where $ b>a>0 $.
	
	\item\label{re3} Subjects do not know  a priori when the words start. Similarly, the classifier has no a priori information about the temporal locations of the words within the noisy sentences.
\end{enumerate}
In the following, we derive three different classifiers implying different assumptions on how humans perform the classification task. We observe, however, that all the classifiers perform almost identically (See Section IV), which means that it does not matter which one we choose to compare with human performance.
\subsubsection{MAP ($ M $)}
\label{subsec:FullBays_Class}
The classifier chooses which word was spoken by maximizing the \textit{posterior probability}: $ P(M=m|\textbf{y}) $; in other words:  
\begin{eqnarray}
\label{Classifier_decision}
m^{*}=\underset{m \in \{1,...,\Gamma\}}{\mathrm{arg} \mathrm{max}} \{P(M=m  |\textbf{y})\}.
\end{eqnarray}
\begin{lemma}
	\label{lemma_Baysian_Classifier}
	The optimal  $ m^{*} $ defined in \eqref{Classifier_decision} is given by:
	\begin{align}
	&m^{*}=\underset{m \in \{ 1,...,\Gamma \}}{\mathrm{arg}\mathrm{max}}\int_{a}^{b}\left(\prod_{z=1}^{v}\dfrac{|2\pi\mathbf{\Sigma}_{\textbf{Y}^{z}|m}|^{-\frac{1}{2}}}{|2\pi\mathbf{\Sigma}_{\textbf{Y}^{z-1}|m}|^{-\frac{1}{2}}}\right)\nonumber\\&\times\exp\frac{-1}{2}\sum_{z=1}^{v}\left((\textbf{y}^{z})^{T}\mathbf{\Sigma}_{\textbf{Y}^{z}|m}^{-1}\textbf{y}^{z}-\left(\textbf{y}^{z-1}\right)^{T}\mathbf{\Sigma}_{\textbf{Y}^{z-1}|m}^{-1}\textbf{y}^{z-1}\right)d\theta,\nonumber
	\end{align}
	\text{where}
	\begin{align}
		&\mathbf{\Sigma}_{\textbf{Y}^{z}|m}=\theta\mathbf{\Sigma}_{\textbf{X}_{m}^{z}}+\mathbf{\Sigma}_{\textbf{W}^{z}},\nonumber\\&\mathbf{\Sigma}_{\textbf{Y}^{z-1}|m}=\theta\mathbf{\Sigma}_{\textbf{X}_{m}^{z-1}}+\mathbf{\Sigma}_{\textbf{W}^{z-1}}.\nonumber
	\end{align}
	\begin{proof}
		See Appendix B.
	\end{proof}
\end{lemma}
\subsubsection{MAP ($ M, \theta \:\text{continuous}$)}
One may argue that subjects are able to identify the SNR  and thereby the scale factor $ \theta $ after having listened to a particular test stimulus,  before deciding on the word. In this case, one should maximize $ f_{M,\Theta|\textbf{y}}(m, \theta |\textbf{y}) $ rather than $ P(M=m |\textbf{y}) $. This leads to the following optimization problem:
\begin{align}
\label{AB_classifier_decision}
(m^{*}, \theta^{*})=\underset{m \in \{ 1,...,\Gamma \}, \theta \in [a, b]}{\mathrm{arg} \mathrm{max}} \{f_{M,\Theta|\textbf{y}}(m, \theta |\textbf{y})\}.
\end{align}
\begin{lemma}
	\label{lemma_ABC_alpha_C}
	The optimal pair $ (m^{*}, \theta^{*}) $ defined in \eqref{AB_classifier_decision} is given by:\footnote{We assume that $a\leq\theta^{*}\leq b $, otherwise the nearest point ($ a $ or $ b $) should be chosen.}
	\begin{align}
	&m^{*}=\underset{m \in \{ 1,...,\Gamma \}}{\mathrm{arg} \mathrm{max}} \left\{ -\sum_{z=1}^{v}\left((\textbf{y}^{z})^{T}(\mathbf{\Sigma}_{\textbf{Y}^{z}|m}^{*})^{-1}\textbf{y}^{z}+\log|\mathbf{\Sigma}_{\textbf{Y}^{z}|m}^{*}|\nonumber\right.\right.\\&\left.\left.-(\textbf{y}^{z-1})^{T}(\mathbf{\Sigma}_{\textbf{Y}^{z-1}|m}^{*})^{-1}\textbf{y}^{z-1}-\log|\mathbf{\Sigma}_{\textbf{Y}^{z-1}|m}^{*}| \right) \right\},\nonumber
	\end{align}
	where 
	\begin{align}
	&\mathbf{\Sigma}_{\textbf{Y}^{z}|m}^{*}=\theta^{*}\mathbf{\Sigma}_{\textbf{X}_{m}^{z}}+\mathbf{\Sigma}_{\textbf{W}^{z}}, \mathbf{\Sigma}_{\textbf{Y}^{z-1}|m}^{*}=\theta^{*}\mathbf{\Sigma}_{\textbf{X}_{m}^{z-1}}+\mathbf{\Sigma}_{\textbf{W}^{z-1}}.\nonumber
	\end{align}
 $ \theta^{*} $ is obtained by solving the following equation with respect to $ \theta $:
\begin{align}
&\sum_{z=1}^{v}\left(-(\textbf{y}^{z})^{T}\left(\mathbf{\Sigma}_{\textbf{Y}^{z}|m}^{-1}\mathbf{\Sigma}_{\textbf{X}_{m}^{z}}\mathbf{\Sigma}_{\textbf{Y}^{z}|m}^{-1}\right)\textbf{y}^{z}+\trace\left(\mathbf{\Sigma}_{\textbf{Y}^{z}|m}^{-1}\mathbf{\Sigma}_{\textbf{X}_{m}^{z}}\right) \nonumber\right.\\&\left.+(\textbf{y}^{z-1})^{T}\left(\mathbf{\Sigma}_{\textbf{Y}^{z-1}|m}^{-1}\mathbf{\Sigma}_{\textbf{X}_{m}^{z-1}}\mathbf{\Sigma}_{\textbf{Y}^{z-1}|m}^{-1}\right)\textbf{y}^{z-1}\nonumber\right.\\&\left.-\trace\left(\mathbf{\Sigma}_{\textbf{Y}^{z-1}|m}^{-1}\mathbf{\Sigma}_{\textbf{X}_{m}^{z-1}}\right)\right)=0.\nonumber
\end{align}
\begin{proof}
See Appendix C.
\end{proof}
\end{lemma}
\subsubsection{MAP ($ M, \theta \:\text{discrete}$)}
\label{subsec:Approxi_Class}
In the version of the listening test used in this paper, a fixed limited set of SNRs are used, and it might be reasonable to assume that the subjects can identify  these SNRs through the training phase. In this case,  the scale factor is a discrete random variable  ($\Theta=\theta_{i}, \quad i \in \{ 1,...,s \} $) rather than a continuous one. Thus,  we maximize $ P(M=m,\Theta=\theta_{i}|\textbf{y}) $. The optimization problem in \eqref{AB_classifier_decision} can thus be rewritten as:
\begin{align}
\label{AB_D_classifier_decision}
(m^{*}, i^{*})=\underset{m \in \{ 1,...,\Gamma \}, i \in \{ 1,...,s \}}{\mathrm{arg} \mathrm{max}} \{P(M=m,\Theta=\theta_{i}|\textbf{y})\}.
\end{align}
\begin{lemma}
	\label{lemma_ABC_D}
	The optimal pair $ (m^{*}, i^{*}) $ defined in \eqref{AB_D_classifier_decision} is given by:
	\begin{align}
	&(m^{*}, i^{*})=\!\!\!\!\!\!\!\!\!\!\!\!\!\!\!\!\underset{m \in \{ 1,...,\Gamma \}, i \in \{ 1,...,s \}}{\mathrm{arg} \mathrm{max}} \!\!\! -\sum_{z=1}^{v}\left((\textbf{y}^{z})^{T}(\mathbf{\Sigma}_{\textbf{Y}^{z}|m}^{i})^{-1}\textbf{y}^{z}\nonumber\right.\\&\left.+\log|\mathbf{\Sigma}_{\textbf{Y}^{z}|m}^{i}|-(\textbf{y}^{z-1})^{T}(\mathbf{\Sigma}_{\textbf{Y}^{z-1}|m}^{i})^{-1}\textbf{y}^{z-1}-\log|\mathbf{\Sigma}_{\textbf{Y}^{z-1}|m}^{i}| \right), \nonumber
	\end{align}
	where 
	\begin{align}
	&\mathbf{\Sigma}_{\textbf{Y}^{z}|m}^{i}=\theta_{i}\mathbf{\Sigma}_{\textbf{X}_{m}^{z}}+\mathbf{\Sigma}_{\textbf{W}^{z}}, \mathbf{\Sigma}_{\textbf{Y}^{z-1}|m}^{i}=\theta_{i}\mathbf{\Sigma}_{\textbf{X}_{m}^{z-1}}+\mathbf{\Sigma}_{\textbf{W}^{z-1}}.\nonumber
	\end{align}
	\begin{proof}
		See Appendix D.
	\end{proof}
\end{lemma}
In order to take into account requirement \ref{re3} that subjects do not know when the word starts a priori,  a window with the same size as the word is shifted within the stimuli. For each shift, the likelihoods  $ P(M=m|\textbf{y}^{w})$ , $  f_{M,\Theta}(m, \theta|\textbf{y}^{w}) $, and $ P(M=m,\Theta=\theta_{i}|\textbf{y}^{w}) $ are calculated using  \cref{lemma_Baysian_Classifier}, \cref{lemma_ABC_alpha_C}, and \cref{lemma_ABC_D}. Denoting by $ \textbf{y}^{w}$ the portion of $ \textbf{y}$ captured by a shift of $ w $, new problems  corresponding to problems \eqref{lemma_Baysian_Classifier}-\eqref{lemma_ABC_D} are respectively formulated as: 
\begin{align}
\label{Classifying_Shifted_loglihood}
&m^{*}=\underset{m \in \{ 1,...,\Gamma \}}{\mathrm{arg} \mathrm{max}} \{\underset{w}{\mathrm{max}} \{ P(M=m|\textbf{y}^{w})\} \},\nonumber\\&(m^{*}, \theta^{*})=\underset{m \in \{ 1,...,\Gamma \}, \theta \in [ a,b]}{\mathrm{arg} \mathrm{max}} \{\underset{w}{\mathrm{max}} \{ f_{M,\Theta|\textbf{y}^{w}}(m, \theta|\textbf{y}^{w})\} \},\nonumber \\&(m^{*}, i^{*})=\!\!\!\!\!\!\!\!\!\!\!\!\!\underset{m \in \{ 1,...,\Gamma \}, i \in \{ 1,...,s \}}{\mathrm{arg} \mathrm{max}}\!\!\!\!\! \{\underset{w}{\mathrm{max}} \{  P(M=m,\Theta=\theta_{i}|\textbf{y}^{w})\} \}.
\end{align}
\section{Simulations And Experiments}
\subsection{Database}
We use the DANTALE II database \cite{wagener2003design} for our simulations. This database contains 150 sentences sampled at  20 kHz and with a resolution of 16 bits. The sentences are spoken by a native Danish speaker. Each sentence is composed of five words from five categories (name, verb, numeral, adjective, object). There are 10 different words in each of the five categories ($ \Gamma=10 $). The sentences are syntactically fixed, but semantically unpredictable (nonsense), i.e. sentences have the same grammatical structure, but do not necessarily make sense. 
\subsection{Listening Test}
We perform a listening test  inspired by the Danish sentence test paradigm DANTALE II \cite{wagener2003design}, which has been designed in order to determine the speech reception threshold (SRT), i.e. the signal-to-noise ratio for which the word recognition rate is 50\%. In our test, the sentences are contaminated with additive stationary  Gaussian noise with the same long-term spectrum as the sentences.  The listening test is composed of two phases: training phase and test phase. In the training phase, we ask normal-hearing subjects to listen to versions of the noisy sentences to familiarize themselves with the test. In the test phase,  subjects listen to the noisy sentences  at different SNRs and they choose the words they hear using a GUI interface. The GUI interface displays all candidate words on a computer screen (i.e. this is a closed-set listening set), and subjects are asked to choose a candidate word for each of the 5 word categories even if they were unable to recognize the words (forced-choice).  In both phases, DANTALE II sentences are used and subjects listen to the noisy sentences using headphones.

\begin{figure}[!]
	\includegraphics[width=\columnwidth]{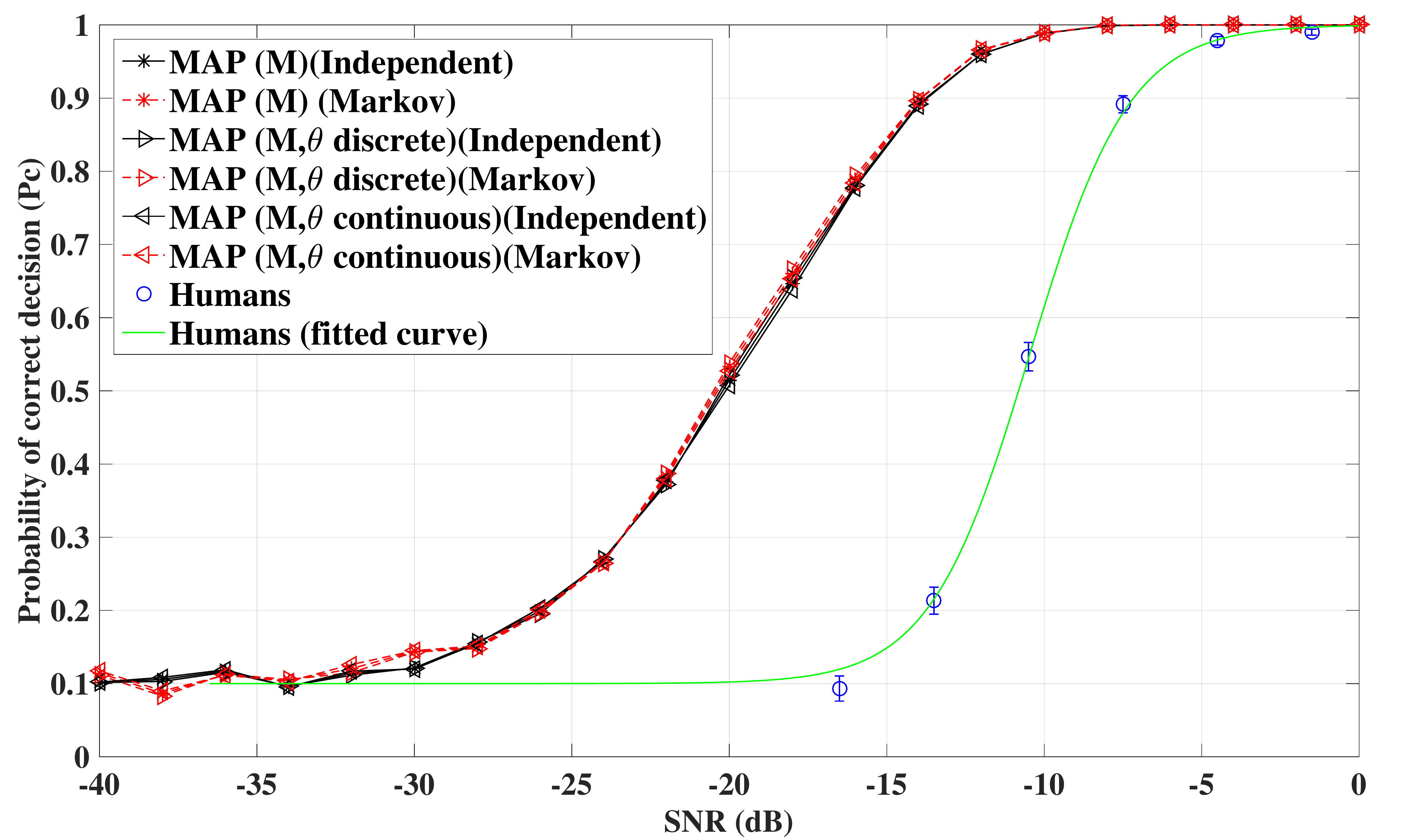}
	\caption{Performance of humans and the optimal classifiers for word recognition as a function of SNR. The error bars show the standard deviation of probability of correct decision among subjects.}
	\label{fig:optimal_classifiers}
\end{figure}

Eighteen normal-hearing native Danish speaking subjects participated in this test. In the training phase, the subjects were exposed to 12 noisy sentences at 6 different SNRs, where each SNR was used twice. In the test phase, each subject listened to 48 sentences (6 SNRs $ \times  $ 8 repetitions). From this listening test, we obtained the human performance for word recognition which is shown in Fig. \ref{fig:optimal_classifiers} (blue circles and green fitted curve). The fitted line is a Maximum Likelihood (ML)-fitted  logistic function of the form $ f(x)=\frac{1-\frac{1}{10}}{1+\exp(cx+d)}+\frac{1}{10} $.
\subsection{Computing Information and Relative Information Loss}
To calculate $ I_{u}(M;\textbf{Y}) $, and $ I_{l}(M;\textbf{Y}) $ in  \eqref{I_U_L_Extended_Lemma_General}, and performance of  the optimal classifiers, we need to build the covariance matrix $ \mathbf{\Sigma}_{\textbf{Y}_{z}|\textbf{y}^{z-1},m} $ for frames of each word.  In the DANTALE II database,  each word has 15 different realizations ($ \textbf{x}_{m}(j),j\in \{1,2,...,15\} $, where  $ j $ denotes the $ j^{th} $ realization of the $ m^{th} $ word). In our simulations, we use 14 different realizations of words for training (building covariance matrices),  and one realization of the words for testing. Using the leave-one-out method, where 14 realizations
are used for training and the last one for testing, we obtain 15
results whose average is used as the final result. In this way, we assume that  the listeners learn one statistical model (covariance matrices) of sub-words for all realizations of that word through the training phase. To construct the covariance matrix $ \mathbf{\Sigma}_{\textbf{Y}_{z}|\textbf{y}^{z-1},m} $ for each frame, we first build $ \mathbf{\Sigma}_{\textbf{Y}^{z}}=\theta\mathbf{\Sigma}_{\textbf{X}_{m}^{z}}+\mathbf{\Sigma}_{\textbf{W}^{z}} $. We  segment each word into $ v $ non-overlapping frames with a duration of 20 ms. We then stack the same sequence of 14 realizations in a long vector $ [\textbf{x}_{m}^{z}(1),\textbf{x}_{m}^{z}(2),...,\textbf{x}_{m}^{z}(14)] $. Then the vector of the linear prediction (LP) coefficients ($ \textbf{a}_{\textbf{X}_{m}^{z}} $) of this long vector is obtained, and the covariance matrix $ \mathbf{\Sigma}_{\textbf{X}_{m}^{z}} $ of this sequence is calculated as described in \cite{godsill1993restoration}. In a similar manner, we construct $ \mathbf{\Sigma}_{\textbf{W}^{z}}  $ using the LP coefficient ($ \textbf{a}_{\textbf{W}^{z}} $) of a long vector that is built by stacking all realizations of all words. The covariance matrices of   $ \mathbf{\Sigma}_{\textbf{Y}_{z}|m} $ and $ \mathbf{\Sigma}_{\textbf{Y}_{z}\textbf{Y}^{z-1}|m} $ are the sub-matrices of $ \mathbf{\Sigma}_{\textbf{Y}^{z}|m} $. Therefore, using \eqref{Covariance_Matrix},  $ \mathbf{\Sigma}_{\textbf{Y}_{z}|\textbf{y}^{z-1},m} $ can be calculated. In our simulations, we consider two cases. In the first case, we assume that frames are independent of each other $ f_{\textbf{Y}_{z}|\textbf{y}^{z-1}}(\textbf{y}_{z}|\textbf{y}^{z-1})=f_{\textbf{Y}_{z}}(\textbf{y}_{z}) $. In the second case, we consider a Markov model of first order $ f_{\textbf{Y}_{z}|\textbf{y}^{z-1}}(\textbf{y}_{z}|\textbf{y}^{z-1})=f_{\textbf{Y}_{z}|\textbf{y}_{z-1}}(\textbf{y}_{z}|\textbf{y}_{z-1}) $.

\begin{figure}[!]
	\includegraphics[width=\columnwidth]{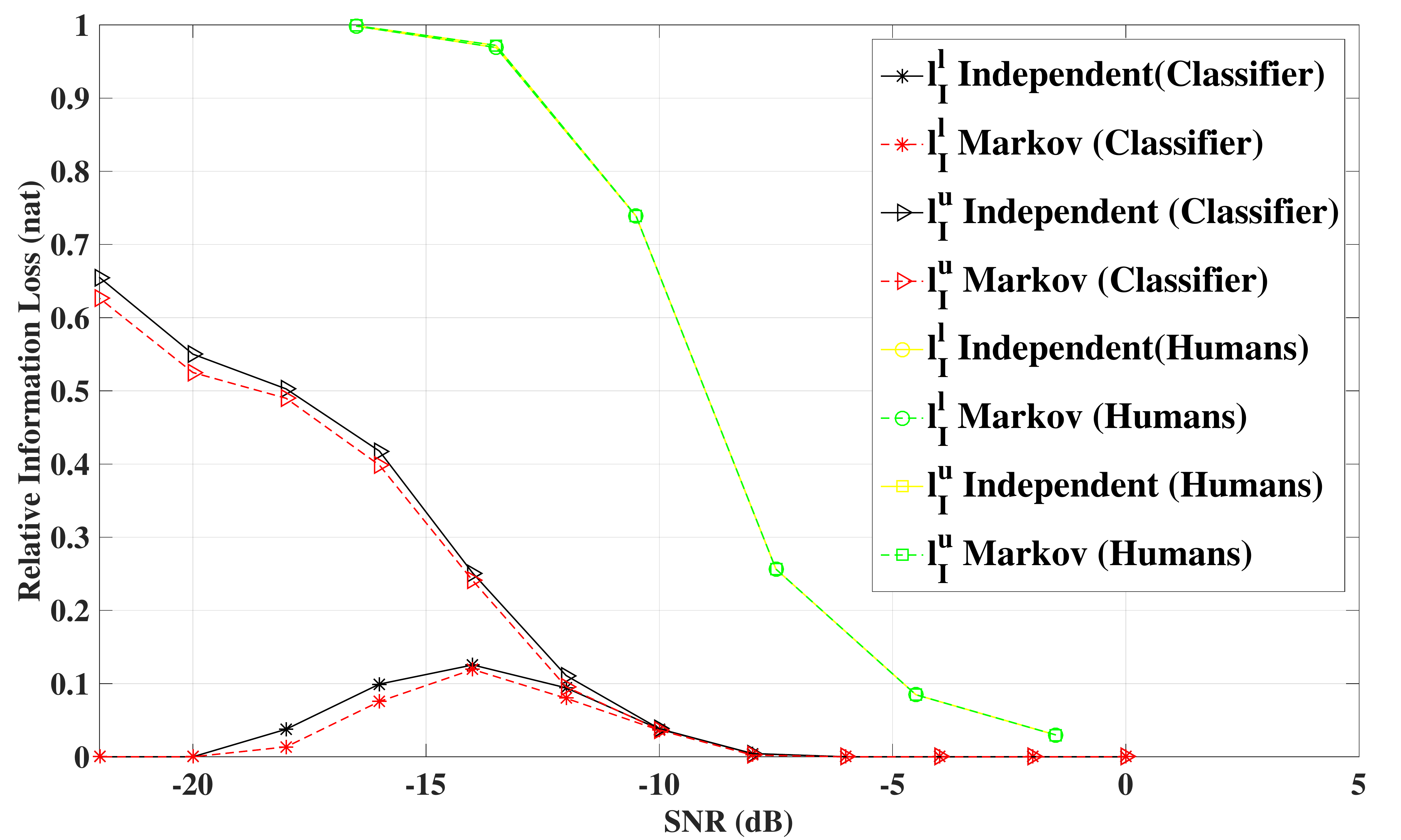}
	\caption{Upper and lower bounds for the relative information loss $ l_{I} $ in the human auditory system and the relative information loss in the optimal classifier as a function of SNR. }
	\label{fig:Information_Loss}
\end{figure}
\begin{figure}[!]
	\includegraphics[width=\columnwidth]{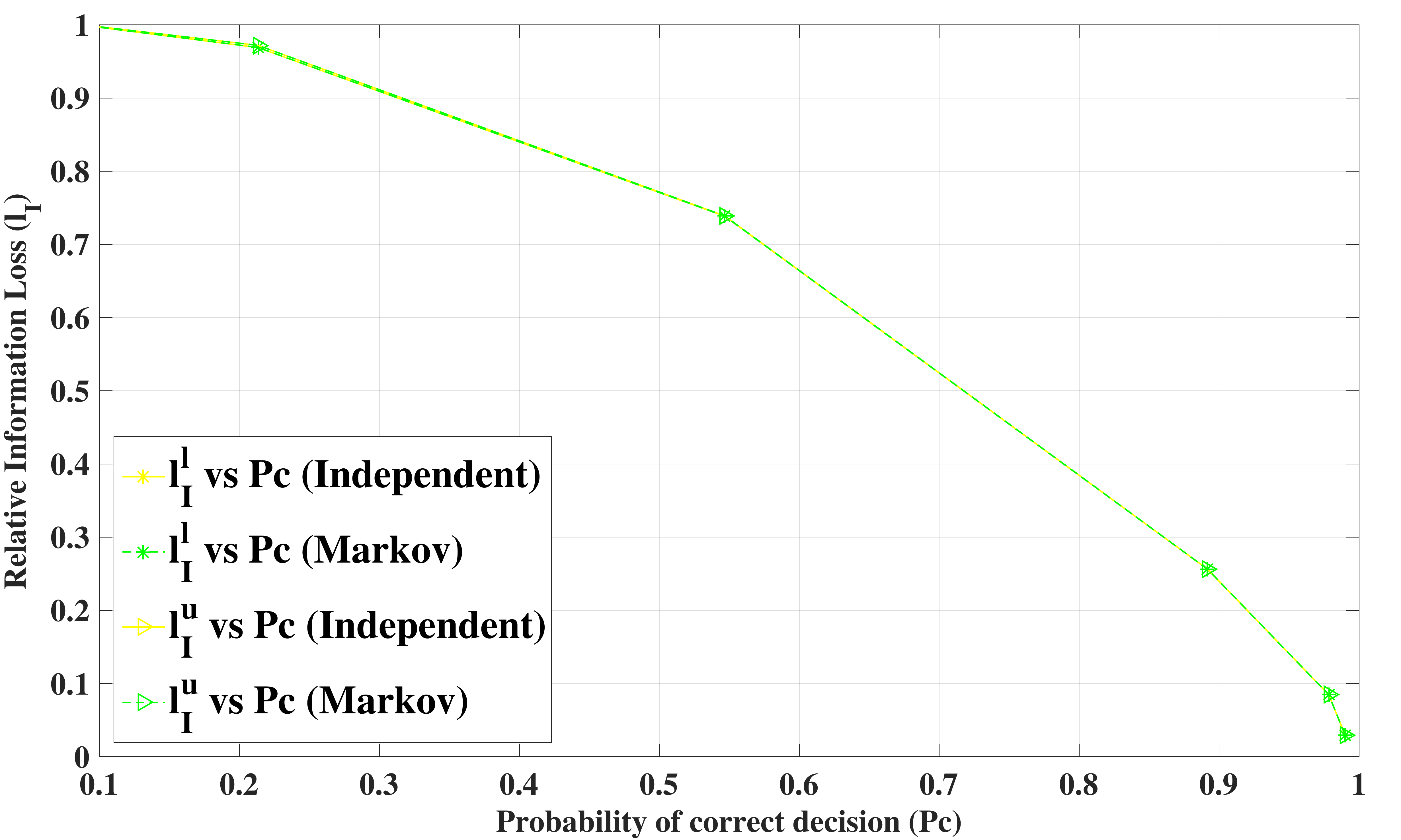}
	\caption{The relative information loss $ l_{I} $ vs. probability of correct decision, $ P_{c} $, for humans.}
	\label{In_Loss_Vs_PD}
\end{figure}

 Using \eqref{adel} and \eqref{I_U_L_Extended_Lemma_General}, we calculate the bounds on the relative information loss in the human auditory system and the relative information loss in the optimal classifiers. The result is plotted in Fig. \ref{fig:Information_Loss}. The relationship between the relative information loss and probability of detection ($ P_{c} $) for humans and  the optimal classifier are  plotted in Fig. \ref{In_Loss_Vs_PD}.


\section{Discussion}

We observe from Fig. \ref{fig:optimal_classifiers} that the word recognition rate for the classifiers and humans reach 1 at high SNRs, whereas at low SNRs, it is at the chance level $ P_{c}=0.1 $. This is because at high noise levels, the words are completely masked by noise, and  both  classifiers  and humans choose words randomly from $ \Gamma=10 $ words. It can also be seen that the optimal classifiers perform almost identically, when we consider the Markov model of the first order compared to the optimal classifiers employing an independent-frame assumption. This implies that the independence assumption does not compromise the performance significantly. This result suggests that an independent-frame assumption, which is often employed in various speech processing contexts (e.g. \cite{deller1993discrete}), is a reasonable assumption, at least in this context. The fact that the performances of all three classifiers are nearly identical,  means that, in this test, the alphabet of the SNR and prior assumptions on it, are insignificant.  Finally, we observe from Fig. \ref{fig:optimal_classifiers} that machine performance is substantially better than human performance. In particular, the $ 50\% $ speech perception threshold for machine receivers is approximately $ 8 $ dB lower than that of humans. The superior performance of classifiers for detecting noisy words compared to humans contradicts our hypothesis  that humans are optimal at recognizing words in noise. In other words, the human auditory system performs sub-optimally in this particular task.


As can be seen from Fig. \ref{fig:Information_Loss}, the relative information loss at high SNRs ($ \text{SNR}>-2\; \text{dB} $) for  humans is around 0, whereas at low SNRs the relative information loss reaches its maximum 1. This is because for SNRs less than  $ -16 $ dB, humans can not recognize the  words and therefore simply guess; so in this case $ I(M;M^{*})=0 $ and $ l_{I}=1 $. We can therefore conclude that not only is less and less information available at the eardrum for decreasing SNRs, but of the information that {\em is} available at the eardrum, less and less is used for identifying the word. On the contrary, at high SNRs, there is no loss of the useful information needed for identifying the words in the human auditory system.  It should be noted that the lower and upper bounds for the relative information loss for humans almost coincide. We also observe that the relative information loss in the optimal classifiers is less than the relative information loss in the human auditory system, which confirms that in this set-up, the human auditory system performs sub-optimally. 

From Fig. \ref{In_Loss_Vs_PD}, it is seen that there is an inverse relationship between probability of correct decision (or  speech intelligibility) and the relative information loss for humans.  This monotonic relationship between intelligibility and the amount of information that reaches the brain has also been observed in \cite{jensen2014speech}.

\section{Conclusion}
In this paper, we defined and quantified the information loss in the human auditory system.  We first considered a speech communication model where words are spoken and sent through a noisy channel, and then received by a listener. For this setup, we defined and bounded the relative information loss in the listener. The relative information loss describes the fraction of speech information that reaches the eardrum of the listener, but which is {\em not} used to decode the speech. To obtain the word recognition rate for humans, we conducted a listening test.  The results showed that bounds for  the relative information loss in the human auditory system are tight and as SNR increases,  the relative information loss decreases. We also assessed the hypothesis that whether humans are optimal in recognizing speech signals in noise. To do so, we derived optimal classifiers and compared their performance for information loss and word recognition rate to those of humans. The lower information loss and higher word recognition rate for machine classifiers compared to humans implied the sub-optimality of the human auditory system for recognizing noisy words, at least for speech contaminated by additive, Gaussian, speech-shaped noise.


%

\appendices
\section{Proof of Lemma \ref{lemma_Up_Low_In_General}}
Upper and lower bounds for the entropy of  the mixture of distributions $ f_{\textbf{Y}}(\textbf{y})=\sum_{m=1}^{\Gamma}P(M=m)f_{\textbf{Y}|m}(\textbf{y}|m) $ are obtained as \cite{kolchinsky2017estimating}:
\begin{align}
\label{Upper_Lower_Entropy}
	&h_{l}(\textbf{Y})=h(\textbf{Y}|M)-\nonumber\sum_{m=1}^{\Gamma}P(M=m)\\\times&\log\left(\sum_{m'=1}^{\Gamma}P(M=m')e^{-C_{\beta}(f_{\textbf{Y}|m}(\textbf{y}|m)||f_{\textbf{Y}|m'}(\textbf{y}|m'))}\right),\nonumber\\&h_{u}(\textbf{Y})=h(\textbf{Y}|M)-\nonumber\sum_{m=1}^{\Gamma}P(M=m)\\\times&\log\left(\sum_{m'=1}^{\Gamma}P(M=m')e^{-D_{KL}(f_{\textbf{Y}|m}(\textbf{y}|m)||f_{\textbf{Y}|m'}(\textbf{y}|m'))}\right),
\end{align}
where $ h_{u}(\textbf{Y}) $ and $ h_{l}(\textbf{Y}) $ denote the upper and lower bounds for $ h(\textbf{Y}) $ respectively. From \eqref{chain_rule}, we have:
\begin{align}
\label{F_Y_z_alpha}
f_{\textbf{Y}|m}(\textbf{y})=&\prod_{z=1}^{v}f_{\textbf{Y}_{z}|\textbf{y}^{z-1},m}(\textbf{y}_{z}|\textbf{y}^{z-1},m)
\end{align}
So the KL-divergence and the Chernoff $ \beta $-divergence in \eqref{Upper_Lower_Entropy} can be written as:
\begin{align}
\label{Sum_Divergence}
&D_{KL}(f_{\textbf{Y}|m}||f_{\textbf{Y}|m'})\nonumber\\&=\sum_{z=1}^{v}E_{\textbf{y}^{z-1}}\left[D_{KL}(f_{\textbf{Y}_{z}|\textbf{y}^{z-1},m}||f_{\textbf{Y}_{z}|\textbf{y}^{z-1},m'})\right],\nonumber\\&C_{\beta}(f_{\textbf{Y}|m}||f_{\textbf{Y}|m'})\nonumber\\&=C_{\beta}(f_{\textbf{Y}^{v-1}|m}f_{\textbf{Y}_{v}|\textbf{y}^{v-1},m}||f_{\textbf{Y}^{v-1}|m'}f_{\textbf{Y}_{v}|\textbf{y}^{v-1},m'}).
\end{align}

 Using \eqref{Upper_Lower_Entropy} and the fact that $ 0\le I(M;\textbf{Y})\le H(M)=\log(\Gamma) $, we find a  lower and upper bound for $ I(M;\textbf{Y}) $:
\begin{align}
\label{Initial_Upper_Lower}
&I_{l}(M;\textbf{Y})=\max\left\{0,h_{l}(\textbf{Y})-h(\textbf{Y}|M)\right\},\nonumber\\&I_{u}(M;\textbf{Y})=\min\left\{\log(\Gamma),h_{u}(\textbf{Y})-h(\textbf{Y}|M)\right\}.
\end{align}
Using the result in \eqref{Sum_Divergence} and substituting \eqref{Upper_Lower_Entropy} in \eqref{Initial_Upper_Lower}  completes our proof.
\section{Proof of lemma \ref{lemma_Baysian_Classifier}}
Using  Bayes' theorem, the \textit{posterior probability} can be written as \cite{proakisdigital}: 
\begin{equation}
\label{Baye_theory_Baysian_classifier}
P(M=m|\textbf{y})= \dfrac{f_{\textbf{Y}|m}(\textbf{y}|m)P(M=m)}{f_{\textbf{Y}}(\textbf{y})}. 
\end{equation}
 Since the speaker chooses words uniformly, $ P(M=m)=\frac{1}{\Gamma},\forall m $, and $ f_{\textbf{Y}}(\textbf{y}) $ is independent of $ M $, from \eqref{Baye_theory_Baysian_classifier}, \eqref{Classifier_decision} can be rewritten:
\begin{align}
\label{Classifier_Decision_Extended}
m^{*}=&\underset{m \in \{ 1,...,\Gamma \}}{\mathrm{arg} \mathrm{max}} \{P(M=m  |\textbf{y})\}=\nonumber\\&\underset{m \in \{ 1,...,\Gamma \}}{\mathrm{arg} \mathrm{max}} \{f_{\textbf{Y}|m}(\textbf{y}|m)\}.
\end{align} 
In \eqref{F_Y_z_alpha}, we have obtained the PDF of $ f_{\textbf{Y}|m}(\textbf{y}|m) $. However, the  classifier does not know the SNR, $ \theta $,  so we can write:
\begin{align}
\label{F_Y_Theta_Decoder}
f_{\textbf{Y}|m,\theta}(\textbf{y}|m,\theta)=\prod_{z=1}^{v}\dfrac{\mathcal{N}(\textbf{0},\mathbf{\Sigma}_{\textbf{Y}^{z}|m})}{\mathcal{N}(\textbf{0},\mathbf{\Sigma}_{\textbf{Y}^{z-1}|m})}.
\end{align}
Using \eqref{Covariance_Matrix}, we obtain:
\begin{align}
\label{Covarince_Matrix_Theta}
	&\mathbf{\Sigma}_{\textbf{Y}^{z}|m}=\theta\mathbf{\Sigma}_{\textbf{X}_{m}^{z}}+\mathbf{\Sigma}_{\textbf{W}^{z}},\nonumber\\&\mathbf{\Sigma}_{\textbf{Y}^{z-1}|m}=\theta\mathbf{\Sigma}_{\textbf{X}_{m}^{z-1}}+\mathbf{\Sigma}_{\textbf{W}^{z-1}}.
\end{align}
Using \eqref{F_Y_Theta_Decoder}, we can calculate $ f_{\textbf{Y}|m}(\textbf{y}|m) $ as:
\begin{align}
\label{F_Y_z_alpha_MLD}
&f_{\textbf{Y}|m}(\textbf{y}|m)=\int f_{\textbf{Y}|m,\theta}(\textbf{y}|m,\theta)f_{\Theta}(\theta) d\theta\nonumber\\&=\int \prod_{z=1}^{v}\dfrac{\mathcal{N}(\textbf{0},\mathbf{\Sigma}_{\textbf{Y}^{z}|m})}{\mathcal{N}(\textbf{0},\mathbf{\Sigma}_{\textbf{Y}^{z-1}|m})}\times f_{\Theta}(\theta)d\theta.
\end{align}
Since $ \Theta \sim \mathcal{U}(a, b) $,  we simply get:
\begin{align}
\label{Proof_Baysian}
&m^{*}=\underset{m \in \{1,...,\Gamma \}}{\mathrm{arg} \mathrm{max}} \left\{\int_{a}^{b}\prod_{z=1}^{v}\dfrac{\mathcal{N}(\textbf{0},\mathbf{\Sigma}_{\textbf{Y}^{z}|m})}{\mathcal{N}(\textbf{0},\mathbf{\Sigma}_{\textbf{Y}^{z-1}|m})}\frac{1}{b-a}d\theta\right\}\nonumber\\&=\underset{m \in \{ 1,...,\Gamma \}}{\mathrm{arg}\mathrm{max}}\int_{a}^{b}\left(\prod_{z=1}^{v}\dfrac{|2\pi\mathbf{\Sigma}_{\textbf{Y}^{z}|m}|^{-\frac{1}{2}}}{|2\pi\mathbf{\Sigma}_{\textbf{Y}^{z-1}|m}|^{-\frac{1}{2}}}\right)\nonumber\\&\times\exp\frac{-1}{2}\sum_{z=1}^{v}\left((\textbf{y}^{z})^{T}\mathbf{\Sigma}_{\textbf{Y}^{z}|m}^{-1}\textbf{y}^{z}-\left(\textbf{y}^{z-1}\right)^{T}\mathbf{\Sigma}_{\textbf{Y}^{z-1}|m}^{-1}\textbf{y}^{z-1}\right)d\theta.
\end{align}
\section{Proof of Lemma \ref{lemma_ABC_alpha_C}}
According to Bayes' theorem, $ f_{M,\Theta|\textbf{y}}(m,\theta|\textbf{y}) $ can be written as: 
\begin{equation}
\label{AB_classifier_Baye_theory}
f_{M,\Theta|\textbf{y}}(m,\theta|\textbf{y})= \dfrac{f_{\textbf{Y}|m,\theta}(\textbf{y}|m,\theta)f_{M,\Theta}(m,\theta)}{f_{\textbf{Y}}(\textbf{y})}.
\end{equation}
  Since $ M $ and $ \theta $ are mutually independent, it follows that  $ f_{M,\Theta}(m,\theta)= P(M=m)f_{\Theta}(\theta)=\frac{1}{\Gamma(b-a)}$. Using \eqref{AB_classifier_Baye_theory} and \eqref{F_Y_Theta_Decoder},  \eqref{AB_classifier_decision} can then be expressed as:
\begin{align}
&(m^{*}, \theta^{*})=\underset{m \in \{ 1,...,\Gamma \}, \theta \in [a, b]}{\mathrm{arg} \mathrm{max}} \{f_{\textbf{Y}|m,\theta}(\textbf{y}|m,\theta)\},\nonumber\\& =\underset{m \in \{ 1,...,\Gamma \}, \theta \in [a, b]}{\mathrm{arg} \mathrm{max}}\left(\prod_{z=1}^{v}\dfrac{|2\pi\mathbf{\Sigma}_{\textbf{Y}^{z}|m}|^{-\frac{1}{2}}}{|2\pi\mathbf{\Sigma}_{\textbf{Y}^{z-1}|m}|^{-\frac{1}{2}}}\right)\nonumber\\&\times\exp\frac{-1}{2}\sum_{z=1}^{v}\left((\textbf{y}^{z})^{T}\mathbf{\Sigma}_{\textbf{Y}^{z}|m}^{-1}\textbf{y}^{z}-\left(\textbf{y}^{z-1}\right)^{T}\mathbf{\Sigma}_{\textbf{Y}^{z-1}|m}^{-1}\textbf{y}^{z-1}\right)\nonumber.
\end{align}
By applying the logarithm,  we get : 
\begin{align}
\label{Log_ML_AB_classifier}
&(m^{*}, \theta^{*})=\underset{m \in \{ 1,...,\Gamma\}, \theta \in [a, b]}{\mathrm{arg} \mathrm{max}}  -\sum_{z=1}^{v}\left((\textbf{y}^{z})^{T}\mathbf{\Sigma}_{\textbf{Y}^{z}|m}^{-1}\textbf{y}^{z}\right.\nonumber\\&\left.+\log|\mathbf{\Sigma}_{\textbf{Y}^{z}|m}| -\left(\textbf{y}^{z-1}\right)^{T}\mathbf{\Sigma}_{\textbf{Y}^{z-1}|m}^{-1}\textbf{y}^{z-1}-\log|\mathbf{\Sigma}_{\textbf{Y}^{z-1}|m}|\right) .
\end{align}
 Equation \eqref{Log_ML_AB_classifier}  indicates that the decoder chooses the pair $ (m, \theta)$ maximizing 
 \begin{align}
 \label{g-function}
&g=-\sum_{z=1}^{v}\left((\textbf{y}^{z})^{T}\mathbf{\Sigma}_{\textbf{Y}^{z}|m}^{-1}\textbf{y}^{z}\nonumber+\log|\mathbf{\Sigma}_{\textbf{Y}^{z}|m}| \nonumber \right.\\&\left.-\left(\textbf{y}^{z-1}\right)^{T}\mathbf{\Sigma}_{\textbf{Y}^{z-1}|m}^{-1}\textbf{y}^{z-1}-\log|\mathbf{\Sigma}_{\textbf{Y}^{z-1}|m}|\right).
 \end{align}
 Here $ \textbf{y}^{z} $ and $ \textbf{y}^{z-1} $ are assumed as  constants at the decoder, so using that $ \frac{\partial \log|\mathbf{\Sigma}_{\textbf{Y}^{z}|m}|}{\partial\theta}= \trace\left(\mathbf{\Sigma}_{\textbf{Y}^{z}|m}^{-1}\frac{\partial \mathbf{\Sigma}_{\textbf{Y}^{z}|m} }{\partial\theta}\right)$, and that $ \frac{\partial \mathbf{\Sigma}_{\textbf{Y}^{z}|m}^{-1}}{\partial\theta}= -\mathbf{\Sigma}_{\textbf{Y}^{z}|m}^{-1}\frac{\partial \mathbf{\Sigma}_{\textbf{Y}^{z}|m}}{\partial\theta}\mathbf{\Sigma}_{\textbf{Y}^{z}|m}^{-1}$, we can find $ \theta^{*} $ by taking the derivate of $ g $  with respect to $ \theta $ in the equation \eqref{g-function}:
\begin{align}
\label{Proof_AB_classifier}
& \dfrac{\partial g}{\partial\theta}=\sum_{z=1}^{v}\left(-(\textbf{y}^{z})^{T}\left(\mathbf{\Sigma}_{\textbf{Y}^{z}|m}^{-1}\mathbf{\Sigma}_{\textbf{X}_{m}^{z}}\mathbf{\Sigma}_{\textbf{Y}^{z}|m}^{-1}\right)\textbf{y}^{z}+\trace\left(\mathbf{\Sigma}_{\textbf{Y}^{z}|m}^{-1}\mathbf{\Sigma}_{\textbf{X}_{m}^{z}}\right) \nonumber\right.\\&\left.+(\textbf{y}^{z-1})^{T}\left(\mathbf{\Sigma}_{\textbf{Y}^{z-1}|m}^{-1}\mathbf{\Sigma}_{\textbf{X}_{m}^{z-1}}\mathbf{\Sigma}_{\textbf{Y}^{z-1}|m}^{-1}\right)\textbf{y}^{z-1}\nonumber\right.\\&\left.-\trace\left(\mathbf{\Sigma}_{\textbf{Y}^{z-1}|m}^{-1}\mathbf{\Sigma}_{\textbf{X}_{m}^{z-1}}\right)\right),\nonumber
\end{align}
We obtain  $ \theta^{*} $ by solving $ \dfrac{\partial g}{\partial\theta}=0 $. Finally, $ m^{*} $ is calculated as follows:
\begin{align}
&m^{*}=\underset{m \in \{ 1,...,\Gamma \}}{\mathrm{arg} \mathrm{max}} \left\{ -\sum_{z=1}^{v}\left((\textbf{y}^{z})^{T}(\mathbf{\Sigma}_{\textbf{Y}^{z}|m}^{*})^{-1}\textbf{y}^{z}+\log|\mathbf{\Sigma}_{\textbf{Y}^{z}|m}^{*}|\nonumber\right.\right.\\&\left.\left.-(\textbf{y}^{z-1})^{T}(\mathbf{\Sigma}_{\textbf{Y}^{z-1}|m}^{*})^{-1}\textbf{y}^{z-1}+\log|\mathbf{\Sigma}_{\textbf{Y}^{z-1}|m}^{*}| \right) \right\},\nonumber
\end{align}
where
\begin{align}
&\mathbf{\Sigma}_{\textbf{Y}^{z}|m}^{*}=\theta^{*}\mathbf{\Sigma}_{\textbf{X}_{m}^{z}}+\mathbf{\Sigma}_{\textbf{W}^{z}}, \mathbf{\Sigma}_{\textbf{Y}^{z-1}|m}^{*}=\theta^{*}\mathbf{\Sigma}_{\textbf{X}_{m}^{z-1}}+\mathbf{\Sigma}_{\textbf{W}^{z-1}}\nonumber.
\end{align}
\section{Proof of Lemma \ref{lemma_ABC_D}}
Using  Bayes' theorem, $ P(M=m,\Theta=\theta_{i}|\textbf{y}) $ can be expressed as: 
\begin{equation}
\label{AB_D_classifier_Baye_theory}
P(M=m,\Theta=\theta_{i}|\textbf{y})= \dfrac{f_{\textbf{Y}|m,\theta_{i}}(\textbf{y}|m,\theta_{i})P(M=m,\Theta=\theta_{i})}{f_{\textbf{Y}}(\textbf{y})}. 
\end{equation}
  Because  $ M $ and $ \theta_{i} $ are mutually independent,  $ P(M,\Theta=\theta_{i})= P(M=m)P(\Theta=\theta_{i})=\frac{1}{\Gamma}\times\frac{1}{s}$. Using \eqref{AB_D_classifier_Baye_theory}, \eqref{AB_D_classifier_decision} can be rewritten as:
\begin{align}
\label{ML_AB_D_classifier}
(m^{*}, i^{*})=\underset{m \in \{ 1,...,\Gamma \}, i \in \{ 1,...,s \}}{\mathrm{arg} \mathrm{max}} \{f_{\textbf{Y}|m,\theta_{i}}(\textbf{y}|m,\theta_{i})\}.
\end{align}
Using \eqref{F_Y_Theta_Decoder}, $ f_{\textbf{Y}|m,\theta_{i}} $ is obtained as: 
\begin{align}
&f_{\textbf{Y}|m,\theta_{i}}(\textbf{y}|m,\theta_{i})=\left(\prod_{z=1}^{v}\dfrac{|2\pi\mathbf{\Sigma}_{\textbf{Y}^{z}|m}^{i}|^{-\frac{1}{2}}}{|2\pi\mathbf{\Sigma}_{\textbf{Y}^{z-1}|m}^{i}|^{-\frac{1}{2}}}\right)\nonumber\\&\times\exp\frac{-1}{2}\sum_{z=1}^{v}\left((\textbf{y}^{z})^{T}(\mathbf{\Sigma}_{\textbf{Y}^{z}|m}^{i})^{-1}\textbf{y}^{z}\right.\nonumber\\&\left.-(\textbf{y}^{z-1})^{T}(\mathbf{\Sigma}_{\textbf{Y}^{z-1}|m}^{i})^{-1}\textbf{y}^{z-1}\right),\nonumber
\end{align}
where 
\begin{align}
&\mathbf{\Sigma}_{\textbf{Y}^{z}|m}^{i}=\theta_{i}\mathbf{\Sigma}_{\textbf{X}_{m}^{z}}+\mathbf{\Sigma}_{\textbf{W}^{z}}, \mathbf{\Sigma}_{\textbf{Y}^{z-1}|m}^{i}=\theta_{i}\mathbf{\Sigma}_{\textbf{X}_{m}^{z-1}}+\mathbf{\Sigma}_{\textbf{W}^{z-1}}.\nonumber
\end{align}
By applying the logarithm, the optimal pair $ (m^{*}, i^{*}) $ is obtained as:
\begin{align}
 \label{Proof_AB_D_classifier}
&(m^{*}, i^{*})=\!\!\!\!\!\!\!\!\!\!\!\!\!\!\!\!\underset{m \in \{ 1,...,\Gamma \}, i \in \{ 1,...,s \}}{\mathrm{arg} \mathrm{max}} \!\!\! -\sum_{z=1}^{v}\left((\textbf{y}^{z})^{T}(\mathbf{\Sigma}_{\textbf{Y}^{z}|m}^{i})^{-1}\textbf{y}^{z}\nonumber\right.\\&\left.+\log|\mathbf{\Sigma}_{\textbf{Y}^{z}|m}^{i}|-(\textbf{y}^{z-1})^{T}(\mathbf{\Sigma}_{\textbf{Y}^{z-1}|m}^{i})^{-1}\textbf{y}^{z-1}-\log|\mathbf{\Sigma}_{\textbf{Y}^{z-1}|m}^{i}| \right) \nonumber.
\end{align}

%
%




%
%
%

\bibliographystyle{IEEEtran}
\bibliography{refs}

%

%






\end{document}